\documentclass[twocolumn,superscriptaddress,floatfix,aps,prb,showpacs]{revtex4}
\usepackage{graphicx,subfigure,epsfig}
\usepackage{dcolumn}
\usepackage{amssymb}
\usepackage{times}
\usepackage{amsmath}
\usepackage{amsfonts}
\usepackage{mathrsfs}
\usepackage{setspace}
\usepackage{latexsym}
\usepackage{bbm}
\usepackage{float}
\usepackage{flafter}
\usepackage{bm}
\usepackage{epstopdf}
\usepackage{hyperref}
\usepackage{color}
\usepackage{multirow}

\begin{document}

\title{Anisotropy Driven Transition from Moore-Read State to Quantum Hall Stripes}

\author{Zheng Zhu}
\affiliation{Department of Physics, Massachusetts Institute of Technology, Cambridge, MA, 02139, USA}
\author{Inti Sodemann}
\affiliation{Department of Physics, Massachusetts Institute of Technology, Cambridge, MA, 02139, USA}
\author{D. N. Sheng}
\affiliation{Department of Physics and Astronomy, California State University, Northridge, CA, 91330, USA}
\author{Liang Fu}
\affiliation{Department of Physics, Massachusetts Institute of Technology, Cambridge, MA, 02139, USA}

\begin{abstract}
We investigate the nature of the quantum Hall liquid in a half-filled second Landau level ($n=1$) as a function of band mass anisotropy using numerical exact diagonalization (ED) and density matrix renormalization group (DMRG) methods. We find increasing the mass anisotropy induces a quantum phase transition from the Moore-Read state to a charge density wave state. By analyzing the energy spectrum, guiding center structure factors and by adding weak pinning potentials, we show that this charge density wave is a uni-directional quantum Hall stripe, which has a periodicity of a few magnetic lengths and survives in the thermodynamic limit. We find smooth profiles for the guiding center occupation function that reveal the strong coupling nature of the array of chiral Luttinger liquids residing at the stripe edges.

\end{abstract}

\pacs{73.43.-f, 73.22.Gk, 71.10.Pm}

\maketitle

\emph{Introduction.}---Fractional quantum Hall (FQH) systems ~\cite{Tsui1982,Laughlin83} have proved to be an inexhaustible platform for exotic phases of matter for  more than three decades. In particular, half-filled Landau levels display a rich set of correlated phases depending on the Landau level index. In the lowest Landau level ($n=0$) of Galilean electrons, the composite fermi liquid (CFL) state is realized at half filling~\cite{HLR}. In the second Landau level ($n=1$), it is believed that the Moore-Read (MR) Pfaffian state ~\cite{MR1991}, or its particle-hole conjugate~\cite{Levin,SSLee}, is realized. In higher Landau levels ($n\geq2$) the ground state is believed to be uni-directional charge density waves also known as stripe phases~\cite{HF1,HF2,HF3,stripe99,bubble00,Variation1,Exp1,Exp2,Exp3,Fradkin99,MacDonald00,Barci}.

Most studies of FQH states are concerned with two-dimensional electron systems with spatial rotational symmetry. In recent years, there is a growing interest in exploring FQH states in the absence of full rotational symmetry  ~\cite{Haldane2011,Balram2016,ExpBx,Eisenstein2011,Wang2012,Yang2012,Qiu2012,Papic2013,Yang2013,Mulligan2010,Haldane2016,Johri2016,Fradkin16,Inti2017,Gokmen2010,Xia2011,Jo2017,Ippoliti2017,Maciejko}. In materials with anisotropic band mass tensors, such as AlAs quantum wells~\cite{Shayegan2006,Gokmen2010}, rotational symmetry can be reduced down to its smallest subgroup consisting of the $\pi$ rotation only. An important question is to determine the stability of various FQH states against mass anisotropy. External perturbations such as in-plane fields~\cite{Exp1,ExpBx} or uniaxial strain \cite{Rokhinson2011,Jo2017} also break rotational symmetry and have qualitatively similar effects as mass anisotropy.

The impact of mass anisotropy depends crucially on the nature of the ground state and on the Landau level. In the lowest Landau level, previous numerical studies have demonstrated that  incompressible FQH states are remarkably robust against mass anisotropy~\cite{Wang2012,Yang2012}. The impact of mass anisotropy is expected to be more pronounced in the second Landau level, where numerical studies have indicated that the MR state is close to a charge density wave instability that can be induced by tuning a few Haldane pseudopotentials~\cite{stripe99,bubble00,Rezayi00}. A previous numerical study provided evidence that an explicit mass anisotropy destabilizes the MR state~\cite{Yang2012}, but the nature of the resulting phase was not determined.  A subsequent study~\cite{Papic2013} demonstrated a phase transition from the MR state into a charged ordered phase driven by an in-plane field and argued that the resulting phase would have stripe character.

In addition to these theoretical studies, there is good experimental evidence that a modest in-plane field drives the isotropic incompressible state observed in GaAs at $\nu=5/2$ into a phase with highly anisotropic transport~\cite{Pan1999,Pan2001,Dean2008,Xia2010,Liu2013}, and a recent experiment has even induced a phase transition into an anisotropic phase by tuning isotropic pressure~\cite{Csathy2016}. However, the nature of the resulting anisotropic phase and its connection to the MR state is not fully understood. Earlier experiments suggested a transition from an incompressible state, like the MR, into a compressible state, like the stripe phase, but a more recent experiment has argued for the possibility of transitioning into a highly anisotropic incompressible state~\cite{Liu2013}.

Motivated by these previous studies, in this paper  we perform numerical exact diagonalization (ED) and density matrix renormalization group (DMRG) studies on the half filled $n=1$ LL in the presence of  mass anisotropy. The mass anisotropy has a qualitatively similar effect as the in-plane field~\cite{Yang2012,Papic2013,Kamburov2013,Yang2016}, but it is much simpler to model theoretically. By calculating energy spectra and the static structure factors, we demonstrate that the incompressible MR state transitions into a compressible state with increasing the mass anisotropy. We will provide ample numerical evidence that the resulting compressible state is a unidirectional quantum Hall stripe. In particular, DMRG simulations have allowed us to reach unprecedented system sizes for the stripe state containing as many as $N=36$ electrons, thus providing strong evidence that it remains the true ground state in the thermodynamic limit.

\begin{figure*}[htbp]
\begin{center}
\includegraphics[width=1\textwidth]{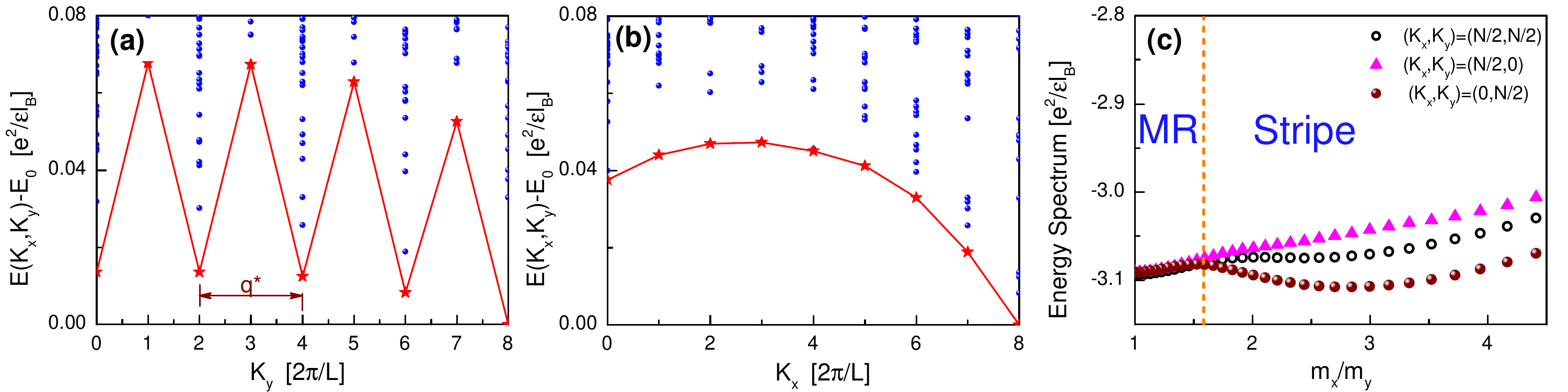}
\end{center}
\par
\renewcommand{\figurename}{Fig.}
\caption{(Color online) The energy spectrum of $N=16$ system got by ED for mass anisotropy $m_y/m_x=0.24$  with momentum $K_y$ (a) and  $K_x$ (b). (c) the MR state evolutes into stripe state with increasing the mass anisotropy, which is characterized by the splitting of the threefold degeneracy of MR state in momentum sectors $(K_x,K_y)=(N/2,N/2),(0,N/2),(N/2,0)$. }
\label{Fig: EK}
\end{figure*}

\emph{Model and Method.}---We consider $N$ spinless electrons moving on a torus subject to a perpendicular magnetic field. The electrons are confined to a Landau level with index $n$ ($n$ LL). We choose Landau gauge $ \mathbf{A}=(0,Bx,0)$ and square geometry in most cases, i.e., $L_x=L_y\equiv L$, satisfying $L_xL_y=2\pi N_\phi$, $N_\phi \in {\mathbb Z}$. Here, $N_\phi$ is the number of flux quanta through the surface. Throughout this paper, we set the magnetic length $l_B\equiv\sqrt{\hbar c/eB}\equiv1$ as  the units of length and $e^2/{\varepsilon l_B}$ as the units of energy.  Then the bare kinetic energy can be written as ${H_0} = \frac{1}{{2m}}{g^{ab}_m}{\Pi _a}{\Pi _b}$ with ${\Pi _a} = {p_a} - \frac{e}{c}{A_a}$ representing the dynamical momentum along $a$ ($a,b=x,y$) direction. Here,   $g_m$ is given by $g_m =\text{diag}[\alpha,1/\alpha]$ in the case of diagonal mass tensor, where $\alpha\equiv\sqrt {m_y/m_x}$ denotes the mass anisotropy, and it determines the shape of LL orbital. Since the kinetic energy of the electrons is quenched due to the magnetic field, the Hamiltonian only includes the projected Coulomb interaction, which reads as
 \begin{equation}\label {Ham}
H_{n \text{LL}}= \frac{1}{N_\phi}\sum\limits_{i < j} {\sum\limits_{{\bf{q}},{\bf{q}} \ne 0} {{V}\left( q_\varepsilon  \right)} } {e^{ - q_m^2/2}}L^2_n(q_m^2/2){e^{i{\bf{q}} \cdot \left( {\bf{R}_{i} - \bf{R}_j} \right)}}.
  \end{equation}

\noindent Here,   $L_n(x)$ is the Laguerre polynomial and $\bf{R}_{i}$ denotes the  guiding center coordinate of the $i$th electron. $V(q_\varepsilon)= 1 /q_\varepsilon$ is the Fourier transform of the Coulomb interaction with $q^2_\varepsilon=g^{ab}_\varepsilon q_aq_b$, where $g^{ab}_\varepsilon$ is the metric derived from the dielectric tensor which defines the shape of the equipotential contours. On the other hand $q^2_m=g^{ab}_mq_aq_b$ includes the metric $g^{ab}_m$ derived from the band mass tensor. Rotational invariance corresponds to congruent metrics, but the physical properties are relevant to their relative difference. Here, we fix $g^{ab}_\varepsilon$ to unity and study mass anisotropy. The impact  of anisotropic dielectric tensor on the Laughlin state has been studied before\cite{Wang2012}.

We use exact diagonalization (ED) and density matrix renormalization group (DMRG) methods. For ED calculation, we take advantage of magnetic translations in two directions in the torus, the states are labeled by the momentum $K=(K_x,K_y)$ in units of $2\pi/L$, and we study systems with up to 16 electrons. For DMRG simulations we impose the conservation of momentum $K_y$ on the torus and keep up to 24000 states, which ensures the truncation error of the order or less than $10^{-6}$, and perform enough number of sweeps (10$\sim$20) to ensure convergence of results.

 \emph{Energy Spectrum.}---We begin by studying the effect of mass anisotropy on the energy spectrum of the half-filled $n=1$ LL. In the isotropic limit, the MR state can be identified as the ground state in our ED calculation, as shown in Fig.~\ref{Fig: EK} (c). The MR state has three topologically degenerate states with momenta $(K_x,K_y)=(N/2,N/2),(0,N/2),(N/2,0)$ in addition to the two-fold center-of-mass degeneracy\cite{Haldane1985,Papic2012} (because of the particle-hole symmetry,  the numerical observed ground states are symmetrized
MR states). Upon introducing mass anisotropy the three-fold degeneracy characteristic of MR state is split beyond a critical value, as shown in Fig.~\ref{Fig: EK} (c), signaling an instability of the MR state into a different phase.   The MR state realized with mass anisotropy can presumably be approximated as the ground state of a three body pseudo-potential model suitably modified to include the nematic distortion of guiding center correlations that can be variationally well approximated by the metric that accounts for these distortions introduced by Haldane~\cite{Haldane2011,Qiu2012}. Figure~\ref{Fig: EK} (a) shows the low energy spectrum of such resulting phase. It is evident that, unlike the MR state, in this phase there is not a recognizable gap separating the ground state manifold from the excited states.  This indicates that the resulting phase with larger mass anisotropy  is compressible. This phase displays a conspicuous set of quasi-degenerate states that differ by a momentum $q^*$. The line that connects the lowest energy states in every momentum sector has a clear zigzag structure as seen in Fig.~\ref{Fig: EK} (a). Interestingly, such zigzag structure only appears in the energy spectrum in one direction, which coincides with the direction of the smaller effective mass. In contrast, Fig.~\ref{Fig: EK} (b)  illustrates the absence of this zigzag structure along the other direction.

The energy spectra provide strong evidence that the resulting state introduced by mass anisotropy is a compressible unidirectional charge density wave. In the next sections we will further study the guiding center form factors and introduce explicit pinning potentials that allow to visualize directly the charge density modulations. This will make clear, beyond any reasonable doubt, that this state is indeed a stripe phase.

 \emph{Structure Factor.}--- To reveal the charge density correlations present in this compressible phase that distinguish it from the MR state and the CFL state in the LLL, we calculate the static structure factor $S_0(q)$ of the density-density correlation, which is defined as
 \begin{equation}
 {S_0}\left( {\bf{q}} \right) = \frac{1}{N}\left\langle {{\rho _{\bf{q}}}{\rho _{ - {\bf{q}}}}} \right\rangle  = \frac{1}{N}\sum\limits_{i,j} {\left\langle {{e^{i{\bf{q}} \cdot {{\bf{R}}_i}}}{e^{ - i{\bf{q}} \cdot {{\bf{R}}_j}}}} \right\rangle }
 \end{equation}
 where ${\rho _{\bf{q}}} = \sum\nolimits_{i = 1}^N {{e^{i{\bf{q}} \cdot {{\bf{R}}_i}}}} $ is the Fourier transform of the guiding center density.

\begin{figure}[htbp]
\begin{center}
\includegraphics[width=0.4\textwidth]{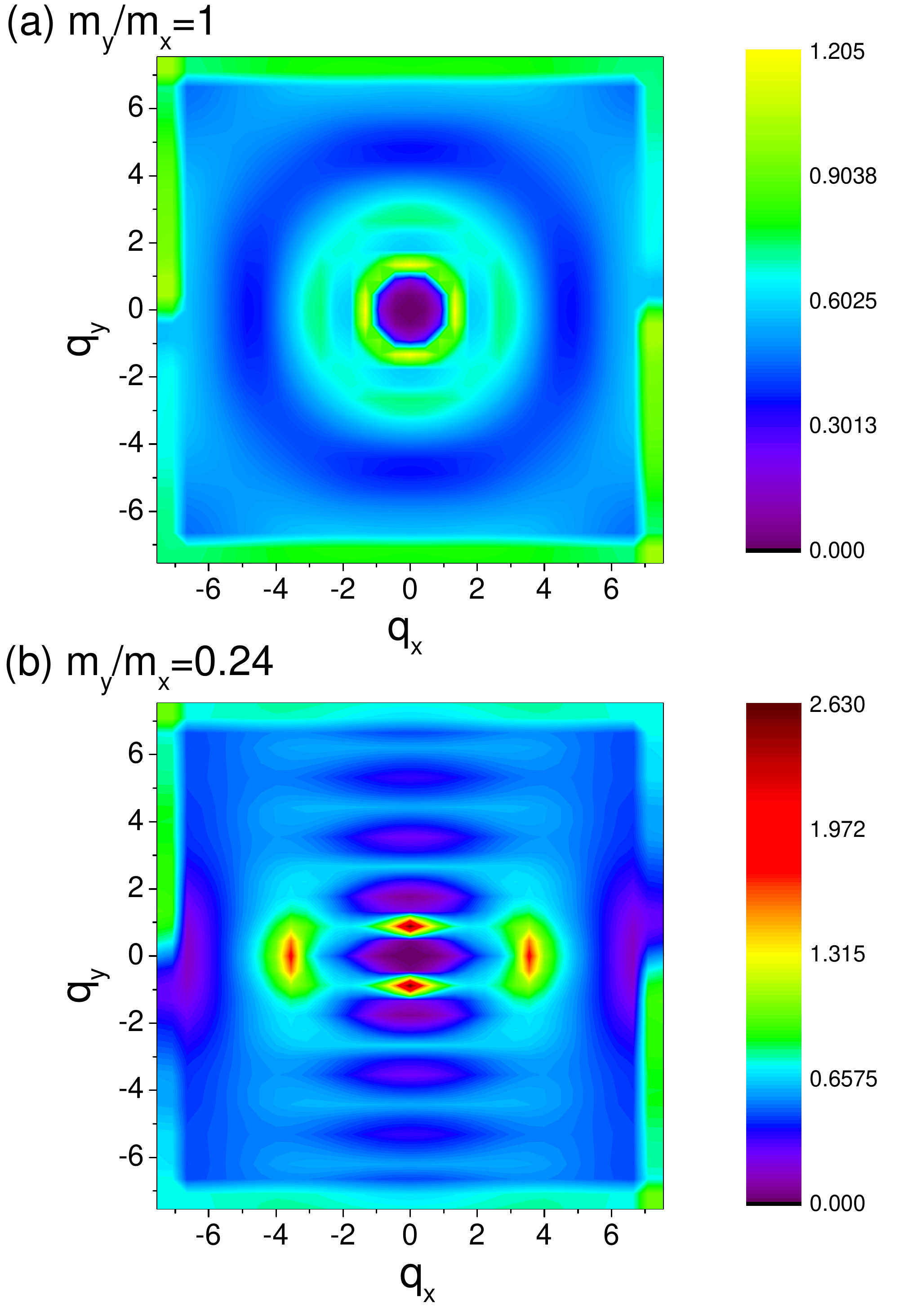}
\end{center}
\par
\renewcommand{\figurename}{Fig.}
\caption{(Color online) The static structure factor of $N=16$ system with the mass ratio (a) $m_y/m_x=1$, (b) $m_y/m_x=0.24$. One can find the well defined sharp peak in the anisotropic case, indicating the charge ordering states arising with mass anisotropy.}
\label{Fig: N16Sq}
\end{figure}

 As shown in Fig.~\ref{Fig: N16Sq} (b), two sharp peaks arise in structure factor $S_0(q)$ when introducing mass anisotropy which are located at $(q_x,q_y)=(0, \pm q^*)$ for $m_y<m_x$. This is in sharp contrast with the MR state realized in the isotropic limit [Fig.~\ref{Fig: N16Sq} (a)]. The existence of peaks in $S_0(q)$  can be regarded as the hallmark of charge ordering. The position of the peaks in $S_0(q)$ represents the wave vector of such charge order, and display stripe features. Here, $q^*$ is exactly the same as the period for the zigzag structure found for the quasi-degenerate states in the energy spectrum [see Fig.~\ref{Fig: EK} (a)], implying the strong density-density correlation in the ground state at this ordering vector. The direction of the charge modulation is the direction with smaller mass, as found in the zigzag structure of energy spectrum.

 \begin{figure*}[htbp]
\begin{center}
\includegraphics[width=1\textwidth]{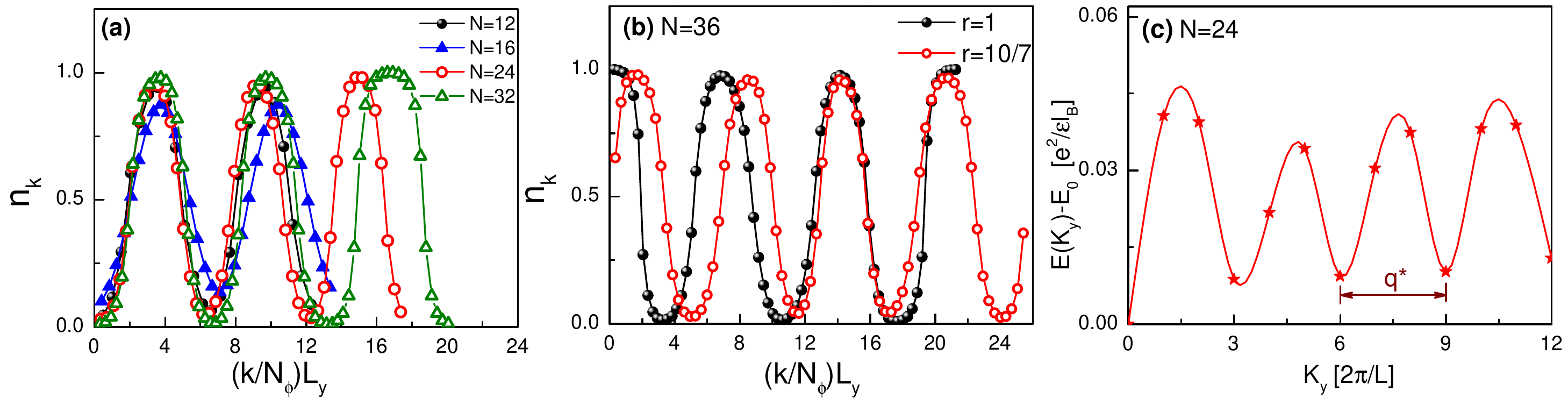}
\end{center}
\par
\renewcommand{\figurename}{Fig.}
\caption{(Color online) The charge density distribution on different orbitals (labeled by $k=1,\ldots, N_\phi$) with adding an onsite potential at one orbital(a) and with different aspect ratios of torus (b). Here, the mass anisotropy ratio is $m_y/m_x=0.24$. (c) The energy spectrum of $N=24$ system obtained by DMRG with mass anisotropy  $m_y/m_x=0.24$.  For the anisotropic limit, one can find the quasi-degenerate states with momentum difference $q^*$  (in the units of $2\pi/L$) exist only in one direction and $q^*$ increases with $L$.}
\label{Fig:ChargeDensity}
\end{figure*}

 \emph{Periodicity of charge modulations.}--- From the above analysis, we have identified an stripe phase that is induced by mass anisotropy and which displays
 instability towards charge modulation along the direction with smaller effective mass. Because the Hamiltonian has translational invariance the ground state cannot break this symmetry explicitly to  display the charge modulations.
Essentially, the ground state of the translationally invariant system is in a Schr\"odinger cat supperposition of charge density wave states that are translated in space relative to each other, and hence its average density is uniform. In order to circumvent this limitation and visualize the charge modulations directly we need to introduce a weak pinning potential. We achieve this task within both the ED and  DMRG numerical simulations which are performed by mapping the single particle orbitals in the Landau gauge into a one dimensional lattice with each site in $y$ direction represents an orbital labeled by momentum in $x$ direction. We add a small on site potential $V_0=0.05$ on one orbital, then we study the charge occupation in each orbital.

As shown in Fig.~\ref{Fig:ChargeDensity}(a), the charge density distribution modulates along different orbitals with amplitude close to 1. Furthermore, One can find the period of  stripe width first increases with $L_y$ as only two periods can  fit in.   Then it jumps to smaller value at $N=24$, where we find three periods. This is consistent with the indication from the energy spectrum, where we find that the value of $q^*$ remains locked at $q^*=2$ (in the unit of $2\pi/L$) for systems for $N=12-16$  electrons, which is the largest sizes within our computational accessibility by ED. However, for a charge density wave state one expects that $q^*$ converges to a non-zero value in the thermodynamic limit ($L\rightarrow \infty$). The observed locking to a specific value in ED can be attributed to the discrete nature of $q^*=2$ which is quantized in units of $2\pi/L$.  To verify this we have performed DMRG simulations on a system with $N=24$. Since $L^2=2\pi N_\phi=4 \pi N$ (at $\nu=1/2$), the system length $L$ is grown by a factor of $\sqrt{24/12}=\sqrt{2}$ comparing to a $N=12$ system. Thus we expect that the $q^*$ obtained at $N=24$ is the closest integer to $2\sqrt{2}(2\pi/L_{N=24})$. In fact, as illustrated in Fig.~\ref{Fig:ChargeDensity}(c), we find that $q^*=3(2\pi/L_{N=24})$ for $N=24$ electrons, in agreement with the expectation that $q^*$ converges to a constant value in the thermodynamic limit.

We find the period of the stripe order in the half filled second LL with anisotropy is about 5 to 6 magnetic lengths. We have been able to reach an unprecedented system size in the numerical study of stripe phases of $N=36$ electrons shown in Fig.~\ref{Fig:ChargeDensity}(b), which lend strong support to the stability of the stripe phase in the thermodynamic limit. As shown in Fig.~\ref{Fig:ChargeDensity}(b) three periods of the stripe fit within the system
 with an aspect ratio $r\equiv L_y/L_x=1$.
 By increasing the aspect ratio of the torus more periods can fit into the system, and Fig.~\ref{Fig:ChargeDensity}(b) shows an example in which the number of periods has been increased to four in this fashion.
For these large system simulations, the density modulation appears in the groundstate even without the pinning force
as the DMRG automatically selects minimum entangled state instead of the  cat state due to the finite trucation error of  the order of $10^{-6}$.

 \emph{Discussion.}---We begin our discussion by explaining the orientation of the stripes relative to the mass tensor. In the presence of mass anisotropy, the single-particle wavefunctions become more localized along the direction of larger mass and more extended along the direction of smaller mass. The orientation of stripes results from a competition between Hartree and exchange energies. The Hartree energy cost, which results from the electrostatic energy associated with charge density modulations, is reduced when the amplitude of charge modulations is reduced. Since the amplitude of charge density modulations is smaller with more delocalized wavefunctions, the Hartree energy cost is reduced when the
charge density modulations occur along the direction with smaller mass. On the other hand, the exchange energy gain tends to increase when the amplitude of the charge density modulations increases, because in this way the electrons tend to be closer to each other in the high density regions of the stripe and therefore gain more energy from the Pauli exclusion principle. Therefore the exchange energy gain is enhanced when direction of charge modulations coincide with the direction of the larger mass. According to these ideas, our results indicate that it is the electrostatic Hartree energy which is dominant in dictating the orientation of the stripes, which is reasonable given that we are using the bare Coulomb interaction. We comment in passing that Hartree-Fock studies have explored the impact of in-plane fields in quantum wells with finite width~\cite{Jungwirth,Phillips}, but recent experiments have found intriguing re-orientations of the stripes as a function of the strength of in-plane field~\cite{Shi1,Shi2} which do not completely fit within the expectations of these earlier theories.

We would like to note, however, that the stripes we observe possibly have a substantial contribution to their energy from correlations beyond those of the Hartree-Fock separation into Hartree and exchange terms. This is evident from the occupation of the guiding center orbitals that is illustrated in Fig.~\ref{Fig:ChargeDensity}, which is in stark contrast with the Hartree-Fock expectation of this being piecewise periodic function that alternates between $0$ (completely empty) and $1$ (completely occupied). The fact that the occupation changes smoothly as opposed to jumping discontinuously between $0$ and $1$ can be viewed as a collapse of the electron quasiparticle residue of the chiral mode that resides at the interface between adjacent electron and hole strips. This behavior is naturally expected within the picture of stripes as an array of coupled Luttinger liquids~\cite{MacDonald00,Barci}.

Since our Hamiltonian explicitly breaks rotational symmetry, the stripe phase we realize only breaks spontaneously the translational symmetry. This feature stabilizes the stripe phase against thermal fluctuations, and, we expect that, in the absence of disorder, there should be power law quasi-long range order for temperatures below a Kosterlitz-Thouless phase transition~\cite{XY}. However, the disorder potential is expected to couple in analogous fashion as a random field to an XY model and, following the Imry-Ma argument, one expects it to destroy the long-range translational order~\cite{ImryMa,Cardy}.

We have found a critical mass anisotropy of $m_x/m_y\approx 1.5$ for the transition from MR into stripes. We wish to emphasize that this value is small compared to the one realized in AlAs quantum wells~\cite{Gokmen2010}, for which $m_x/m_y\approx 5$. Therefore, we expect that when a single valley is occupied in the $n=1$ LL of AlAs the ground state of the system is in the stripe phase. Valley polarization in AlAs can be enforced by applying a modest amount strain~\cite{Gokmen2010}.

In summary we have identified how the MR state, realized in isotropic half-filled $n=1$ LL, undergoes a phase transition into a unidirectional charge density wave state with increasing mass anisotropy. This is shown by the splitting of the topological degeneracy of the MR state in Fig.~\ref{Fig: EK} (c) beyond a critical mass anisotropy. We have performed various tests that demonstrate that the resulting phase has uni-directional translational broken symmetry, including the analysis of the spectrum, guiding center structure factors, and by adding explicit weak pinning potentials that allow to visualize directly the modulations of the occupation of the guiding center orbitals. The charge density modulations of the stripes are found to take place along the direction with smaller mass.

\begin{acknowledgments}
We would like to thank Michael Zudov, Qianhui Shi, Shafayat Hossain, Mansour Shayegan, Eduardo Fradkin, Roderich Moessner and Zlatko Papi\'{c} for insightful discussions. Z.Z.  and L.F. are supported by the David and Lucile Packard foundation. I.S. is supported by the Pappalardo Fellowship.  D.N. Sheng is supported by the U.S. Department of Energy, Office of Basic Energy Sciences under grants No.  DE-FG02-06ER46305.  D.N. Sheng was supported in part by the Gordon and Betty Moore Foundation's EPiQS Initiative, grant GBMF4303 during her visit at MIT.  Z.Z. used the Extreme Science and Engineering Discovery Environment (XSEDE) to perform part of simulation, which is supported by National Science Foundation grant number ACI-1053575.
\end{acknowledgments}


\begin{thebibliography}{99}
\bibitem{Tsui1982} D. C. Tsui, H. L. Stormer, and A. C. Gossard, Phys. Rev. Lett. 48, 1559 (1982).
\bibitem{Laughlin83} R. B. Laughlin, Phys. Rev. Lett. 50, 1395 (1983).
\bibitem{HLR} B. I. Halperin, P. A. Lee, and N. Read, Phys. Rev. B 47, 7312 (1993).
\bibitem{MR1991}  G. Moore and N. Read, Nucl. Phys. B \textbf{360}, 362 (1991).
\bibitem{Levin} M. Levin, B. I. Halperin, and B. Rosenow, Phys. Rev. Lett. 99, 236806 (2007).
\bibitem{SSLee} Sung-Sik Lee, Shinsei Ryu, Chetan Nayak, and Matthew P. A. Fisher, Phys. Rev. Lett. 99, 236807 (2007).

\bibitem{HF1}A. A. Koulakov, M. M. Fogler, and B. I. Shklovskii, Phys. Rev. Lett. \textbf{76}, 499 (1996).
\bibitem{HF2}M. M. Fogler, A. A. Koulakov, and B. I. Shklovskii, Phys. Rev. B \textbf{54}, 1853 (1996).
\bibitem{HF3}R. Moessner and J. T. Chalker, Phys. Rev. B \textbf{54}, 5006 (1996).

\bibitem{Exp1}M. P. Lilly, K. B. Cooper, J. P. Eisenstein, L. N. Pfeiffer, and K. W. West, Phys. Rev. Lett. {\bf 82}, 394 (1999).
\bibitem{Variation1}M. M. Fogler and A. A. Koulakov, Phys. Rev. B \textbf{55}, 9326 (1997).
\bibitem{stripe99}E. H. Rezayi, F. D. M. Haldane, and K. Yang, Phys. Rev. Lett. \textbf{83}, 1219 (1999).
\bibitem{bubble00}F. D. M. Haldane, E. H. Rezayi, and Kun Yang, Phys. Rev. Lett. \textbf{85}, 5396 (2000).
\bibitem{Exp2} R. R. Du, D. C. Tsui, H. L. Stormer, L. N. Pfeiffer, K. W. Baldwin, and K. W. West, Solid State Commun. 109, 389 (1999).
\bibitem{Exp3} K. B. Cooper, M. P. Lilly, J. P. Eisenstein, L. N. Pfeiffer, and K. W. West, Phys. Rev. B 65, 241313(R) (2002).
\bibitem{Fradkin99} E. Fradkin and S. A. Kivelson, Phys. Rev. B 59, 8065 (1999); E. Fradkin, S. A. Kivelson, E. Manousakis, and K. Nho, Phys. Rev. Lett. 84, 1982 (2000).
\bibitem{MacDonald00}A. H. MacDonald and M. P. A. Fisher, Phys. Rev. B 61, 5724 (2000).
\bibitem{Barci} D. G. Barci, E. Fradkin, S. A. Kivelson, and V. Oganesyan, Phys. Rev. B {\bf 65}, 245319 (2002).

\bibitem{Haldane2011}F. D. M. Haldane, Phys. Rev. Lett. 107, 116801 (2011).

\bibitem{Mulligan2010}M. Mulligan, C. Nayak, and S. Kachru, Phys. Rev. B 82, 085102 (2010); Phys. Rev. B 84, 195124 (2011).
\bibitem{Eisenstein2011} J. Xia, J. P. Eisenstein, L. N. Pfeiffer, and K. W. West, Nat. Phys. 7, 845 (2011).
\bibitem{Wang2012} H. Wang, R. Narayanan, X. Wan, and F. Zhang, Phys. Rev. B {\bf 86}, 035122 (2012).
\bibitem{Yang2012}Bo Yang, Z. Papi\'{c}, E. H. Rezayi, R. N. Bhatt, and F. D. M. Haldane, Phys. Rev. B \textbf{85}, 165318 (2012).
\bibitem{Qiu2012}R. Z. Qiu, F. D. M. Haldane, X. Wan, K. Yang, and S. Yi, Phys. Rev. B 85, 115308 (2012).
\bibitem{Papic2013}Z. Papi\'{c}, Phys. Rev. B 87, 245315 (2013).
\bibitem{Yang2013}K. Yang, Phys. Rev. B 88, 241105(R) (2013).
\bibitem{ExpBx} Y. Liu, S. Hasdemir, M. Shayegan, L. N. Pfeiffer, K. W. West, and K. W. Baldwin, Phys. Rev. B 88, 035307 (2013); D. Kamburov, Y. Liu, M. Shayegan, L. N. Pfeiffer, K. W. West, and K. W. Baldwin, Phys. Rev. Lett. 110, 206801(2013); D. Kamburov, M. A. Mueed, M. Shayegan, L. N.Pfeiffer, K. W. West, K. W. Baldwin, J. J. D. Lee, and R. Winkler, Phys. Rev. B 89, 085304 (2014); M. A. Mueed, D. Kamburov, Y. Liu, M. Shayegan, L. N. Pfeiffer, K. W. West, K. W. Baldwin, and R. Winkler, Phys.Rev. Lett. 114, 176805 (2015);M. A. Mueed, D. Kamburov, S. Hasdemir, L. N. Pfeiffer, K. W. West, K. W. Baldwin, and M. Shayegan,Phys. Rev. B 93, 195436(2016).
\bibitem{Haldane2016}F. D. M. Haldane and Y. Shen, arXiv:1512.04502 (2016).
\bibitem{Johri2016}S. Johri, Z. Papic, P. Schmittikert, R. N. Bhatt, and F. D. M. Haldane, New J. Phys 18 (2016).
\bibitem{Balram2016} A. C. Balram and J. K. Jain, Phys. Rev. B {\bf 93}, 075121 (2016).
\bibitem{Fradkin16} Y. You, G. Y. Cho, and E. Fradkin, Phys. Rev. X {\bf 4}, 041050 (2014).
\bibitem{Inti2017}Inti Sodemann, Zheng Zhu, Liang Fu,	arXiv:1701.07836 (2017).

\bibitem{Maciejko}J. Maciejko, B. Hsu, S. A. Kivelson, Y. J. Park, and S. L. Sondhi, Phys. Rev. B \textbf{88}, 125137 (2013);N. Regnault, J. Maciejko, S. A. Kivelson, S. L. Sondhi,arXiv:1607.02178 (2016).




\bibitem{Gokmen2010} T. Gokmen, M. Padmanabhan, and M. Shayegan, Nat. Phys. {\bf 6}, 621 (2010).
\bibitem{Jo2017}Insun Jo, K. A. Villegas Rosales, M. A. Mueed, L. N. Pfeiffer, K. W. West, K. W. Baldwin, R. Winkler, Medini Padmanabhan, M. Shayegan, arXiv: 1701.06684 (2017).
\bibitem{Xia2011} J. Xia, J. P. Eisenstein, L. N. Pfeiffer, and K. W. West, Nat. Phys. 7, 845 (2011).
\bibitem{Ippoliti2017} M. Ippoliti, S. D. Geraedts, and R. N. Bhatt, arXiv: 1701.07832 (2017).

\bibitem{Shayegan2006} M. Shayegan, E. P. De Poortere, O. Gunawan, Y. P. Shkolnikov, E. Tutuc, K. Vakili, Phys. Stat. Sol. b 243, 3629 (2006).

\bibitem{Rokhinson2011} S. P. Koduvayur, Y. Lyanda-Geller, S. Khlebnikov, G. Csathy, M. J. Manfra, L. N. Pfeiffer, K. W. West, and L. P. Rokhinson, Phys. Rev. Lett. {\bf 106}, 016804 (2011).
\bibitem{Rezayi00}E. H. Rezayi and F. D. M. Haldane, Phys. Rev. Lett. \textbf{84}, 4685 (2000).
\bibitem{Jain1989} J. K. Jain, Phys. Rev. Lett. 63, 199 (1989).
\bibitem{Read2000} N. Read and D. Green, Phys. Rev. B \textbf{61}, 10267 (2000).
\bibitem{Pan1999} W. Pan, R. R. Du, H. L. Stormer, D. C. Tsui, L. N. Pfeiffer, K. W. Baldwin, and K. W. West, Phys. Rev. Lett. {\bf 83}, 820 (1999).
\bibitem{Pan2001} W. Pan, J.S. Xia, E.D. Adams, R.R. Du, H.L. Stormer, D.C. Tsui, L. N. Pfeiffer, K.W. Baldwin, and K. W. West, Physica B {\bf 298}, 113 (2001).
\bibitem{Dean2008} C. R. Dean, B. A. Piot, P. Hayden, S. Das Sarma, G. Gervais, L. N. Pfeiffer, and K. W. West, Phys. Rev. Lett. {\bf 101}, 186806 (2008).
\bibitem{Xia2010} J. Xia, V. Cvicek, J. P. Eisenstein, L. N. Pfeiffer, and K. W. West, Phys. Rev. Lett. {\bf 105}, 176807 (2010).
\bibitem{Liu2013} Y. Liu, S. Hasdemir, M. Shayegan, L. N. Pfeiffer, K. W. West, and K. W. Baldwin, Phys. Rev. B {\bf 88}, 035307(2013).
\bibitem{Csathy2016} N. Samkharadze, K. A. Schreiber, G. C. Gardner, M. J. Manfra, E. Fradkin and G. A. Csáthy, Nature Physics {\bf 12}, 191 (2016).
\bibitem{Kamburov2013} D. Kamburov, M. A. Mueed, M. Shayegan, L. N. Pfeiffer, K. W. West, K. W. Baldwin, J. J. D. Lee, and R. Winkler, Phys. Rev. B {\bf 88}, 125435 (2013).
\bibitem{Yang2016} Bo Yang, Zi-Xiang Hu, Ching Hua Lee, Zlatko Papi\'{c}, arXiv:1609.06730 (2016).

\bibitem{Haldane1985} F. D. M. Haldane, Phys. Rev. Lett. \textbf{55}, 2095 (1985).
\bibitem{Papic2012} Z. Papi\'c, F. D. M. Haldane, and E. H. Rezayi, Phys. Rev. Lett. {\bf 109}, 266806 (2012).

\bibitem{Jungwirth} T. Jungwirth, A. H. MacDonald, L. Smr$\breve{c}$ka, and S. M. Girvin, Phys. Rev. B {\bf 60}, 15574 (1999).
\bibitem{Phillips} T. D. Stanescu, I. Martin, and P. Phillips, Phys. Rev. Lett. {\bf 84}, 1288 (2000).

\bibitem{Shi1} Q. Shi, M. A. Zudov, J. D. Watson, G. C. Gardner, and M. J. Manfra, Phys. Rev. B {\bf 93}, 121411(R) (2016);
\bibitem{Shi2} Q. Shi, M. A. Zudov, J. D. Watson, G. C. Gardner, and M. J. Manfra, Phys. Rev. B {\bf 93}, 121404(R) (2016).

\bibitem{XY} T. Franosch and D. R. Nelson, Phys. Rev. E {\bf 63}, 061706 (2001).

\bibitem{ImryMa} Y. Imry and S-k. Ma, Phys. Rev. Lett. {\bf 35}, 1399 (1975).
\bibitem{Cardy} J. L. Cardy and S. Ostlund, Phys. Rev. B {\bf 25}, 6899 (1982).

\end{thebibliography}
\end{document}